# Identification of Cancer: Mesothelioma's Disease Using Logistic Regression and Association Rule

[1,2]Avishek Choudhury

[1]*Systems Engineering, Stevens Institute of Technology, Hoboken, USA*
[2]*Data Science, Syracuse University, Syracuse, USA*

**Abstract:** Malignant Pleural Mesothelioma (MPM) or malignant mesothelioma (MM) is an atypical, aggressive tumor that matures into cancer in the pleura, a stratum of tissue bordering the lungs. Diagnosis of MPM is difficult and it accounts for about seventy-five percent of all mesothelioma diagnosed yearly in the United States of America. Being a fatal disease, early identification of MPM is crucial for patient survival. Our study implements logistic regression and develops association rules to identify early stage symptoms of MM. We retrieved medical reports generated by Dicle University and implemented logistic regression to measure the model accuracy. We conducted (a) logistic correlation, (b) Omnibus test and (c) Hosmer and Lemeshow test for model evaluation. Moreover, we also developed association rules by confidence, rule support, lift, condition support and deployability. Categorical logistic regression increases the training accuracy from 72.30% to 81.40% with a testing accuracy of 63.46%. The study also shows the top 5 symptoms that is mostly likely indicates the presence in MM. This study concludes that using predictive modeling can enhance primary presentation and diagnosis of MM.

## Introduction

Malignant Pleural Mesothelioma (MPM) is a hostile tumor of mesothelial cells concomitant with preceding asbestos contact. With an amplified implementation of chemotherapy (Vogelzang *et al.*, 2003; Zalcman *et al.*, 2016) and a varied gamut of clinical examinations, precise prognostication is a crucial subject for individuals with MPM, doctors and scholars. However, MPM is an outstandingly different ailment. Staging system (Pass *et al.*, 2016), challenging primary tumor identification process (Gill *et al.*, 2016; Frauenfelder *et al.*, 2011) and distinct biology (Bueno *et al.*, 2016), impedes accurate prediction of MM. This fatal disease affects about two individuals per million per annum in a general population (McDonald and McDonald. 1996). Comparatively industrialized nations are affected more by MM (Spirtas *et al.*, 1986; Peto *et al.*, 1995; Leigh *et al.*, 1991) due to higher exposure to asbestos (Metintas *et al.*, 2008). The primitive symptoms of MM such as (a) puffing, (b) dyspnea, (c) respiratory complications, (d) pain in the chest or abdomen, (e) fever and night sweats, (f) pleural effusion, (g) fatigue and (h) muscles weakness does not trigger a doctor to conduct a diagnosis of mesothelioma (MN, 2018) on time.

Predictive analytics can assist in the early discovery of diseases (Choudhury and Greene, 2018; Choudhury and Khan, 2018; Choudhury and Wesabi, 2018). However, distracted symptoms and various malignancies implicating the same tumor or cancer site may lead to a significant fraction of misclassifications leading to poor prediction accuracy. Coalescing evidence from various indicators using data mining techniques, such as Decision Tree (DT), Artificial Neural Network (ANN), Support Vector Machine (SVM) and other classifiers can enhance classification accuracy (Choudhury and Greene, 2018). Regrettably, each method inherits its limitations (Choudhury and Greene, 2018). For example, Random forest, decision tree and tree classifiers either tend to overfit (Choudhury and Khan, 2018) or fail to converge a large dataset (Choudhury and Wesabi, 2018).

In our analysis, we study the classification accuracy of the logistics regression model and compare its testing and training accuracy. Consecutively, our study proves that a logistic regression model gives better prediction than that of a base model.

## Methodology

### Data Description

This study uses the patient's medical reports generated by Dicle University. The dataset contains 34 attributes, one binary response variable and 324 instances. Figure 1 shows the distribution of healthy and mesothelioma patients, where red bars represents mesothelioma patients and blue symbolizes healthy patients.

The dataset consists of 41% females and 59% males. Figure 2 shows the distribution of gender and health outcome.

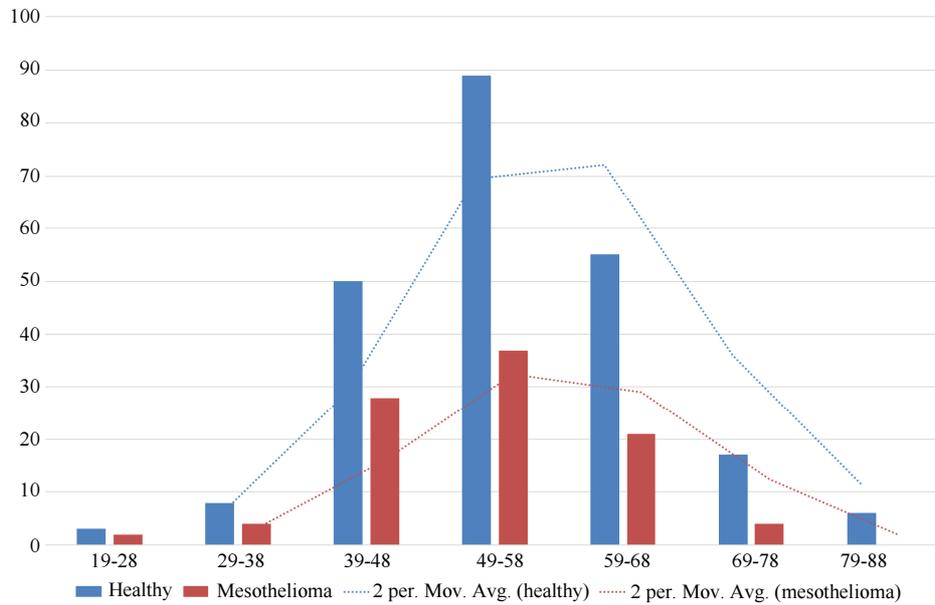

**Fig. 1:** Age versus health outcome

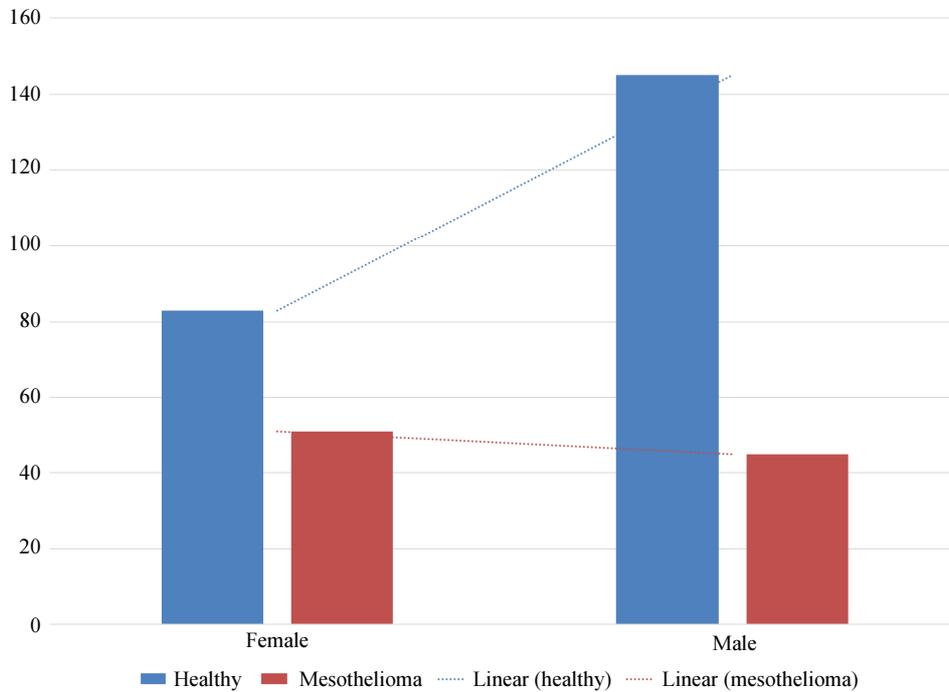

**Fig. 2:** Gender versus health outcome

The patients involved in this study belong to 9 different cities (0 through 8). Figure 3 shows the distribution of patients across cities. The x-axis represents the cities and Y-axis is the count of patients. City 0 consists of 22% healthy (*h*) and 9% mesothelioma (*m*) patients, city 1 consists of 11% *h* and 2% *m*, city 2 consists of 10% *h* and 6% *m*, city 3 consists of 4% *h* and 4% *m*, city 4 consists of 5% *h* and 2% *m*, city 5 consists of 1% *h* and no *m*, city 6 consists of 14% *h* and 6% *m*, city 7 consists of 3% *h* and 1% *m* and 1 individual from city 8 was found to be *h*.

Figure 4 shows the asbestos exposure for each city. It can be observed that the patients from city 0 have the highest asbestos exposure.

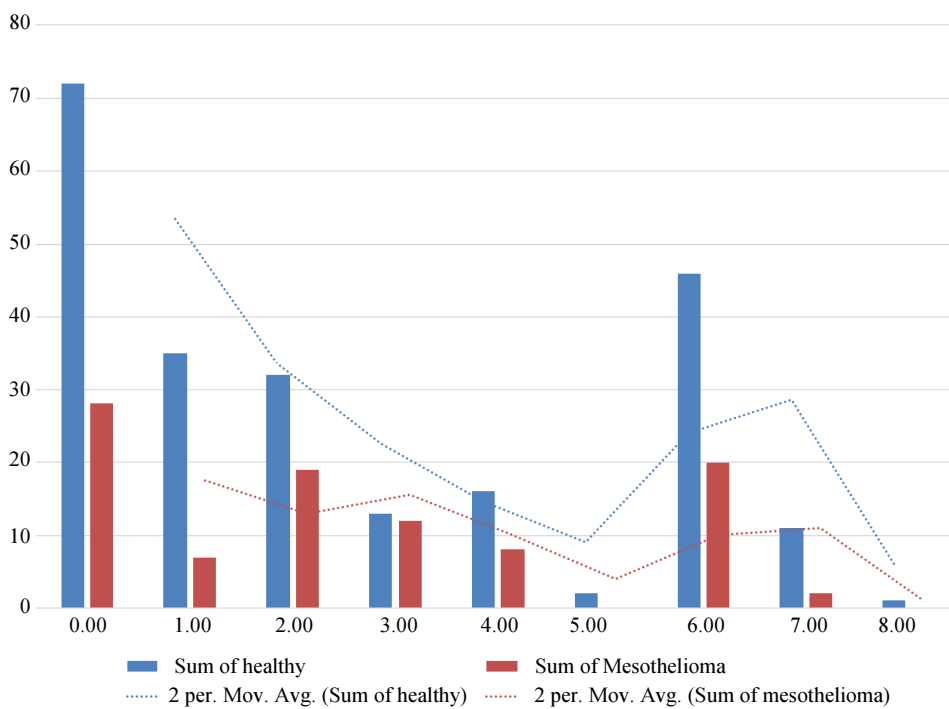

**Fig. 3:** City versus number of patients

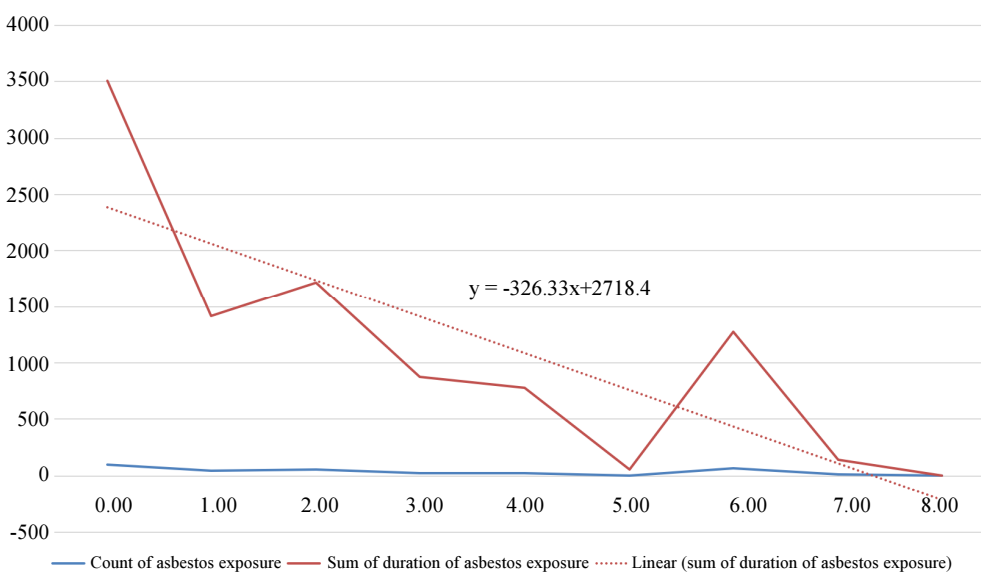

**Fig. 4:** City versus Asbestos exposure

**Table 1:** Data partitioning (training and testing dataset)

| Dataset partitioning summary | | |
|---|---|---|
| | N | Percent |
| Training | 220 | 67.90% |
| Testing | 104 | 32.09% |
| Total | 324 | 100.0% |

*Data Preprocessing*

We partitioned the data into training and testing dataset as shown in Table 1 and used training dataset for all the analysis. The training dataset was balanced using under-sampling (Parodi *et al*., 2015) method.

*Logistic Regression*

Logistic regression is a widely used statistical technique (Tin, 1995). However, its typical use involves situations in which the outcome variable is continuous. Many situations in data analysis involve predicting the value of a nominal or an ordinal categorical outcome variable. In a situation in which we have a nominal categorical outcome variable, we use binary or multinomial logistic regression (Tin, 1995). In this study, we implement binomial logistic regression. Logistic regression is designed to use a mix of continuous and categorical predictor variables to predict a nominal categorical dependent variable. It does not directly predict the values of the dependent variable. Instead, the logistic equation predicts the odds of the event of interest occurring.

The general equation for logistic regression is:

$$Ln(Odds) = \alpha + B_1 X_1 + B_2 X_2 + \ldots + B_k X_k \qquad (1)$$

where, the terms on the right are the standard terms for the independent variable and the intercept in a regression equation (Tin, 1995), on the left side of the equation is the natural log of the odds (ln (Odds)) known as logit (Tin, 1995). The logit function is an S-shaped function. Logistic regression is managed by learning from the function as $P(y = x)$. *Y* is a discrete value and *X* is a vector including discrete or continuous values. The algorithm estimates parameters from the training dataset. Logistic regression algorithm concludes probability and classifies the testing value by using threshold. Post optimizing the equation parameters, it can be employed to predict the output of testing dataset (Choudhury and Wesabi, 2018). Our study employs Categorical Regression which extends the regression model by quantifying categorical variables (Tin, 1995). It can also reduce multicollinearity among predictors, can model nonlinear relationships (Tin, 1995). Categorical regression maximizes the squared correlation between the transformed dependent variable and the linear combination of the transformed predictors (Tin, 1995).

Categorical regression models make the same assumptions as linear regression models. Besides, categorical regression models assume that:

- There cannot be negative numbers in the data and all values must be integers (decimal digits are truncated)
- All nominal and ordinal variables should be coded so that their values are consecutive integers beginning with 1

In classification applications, calculating logistic dependencies between a single input and single target or class variable can be helpful. It determines the outright values of the logistic correlation concerning all predictors and all response variables. The logistic correlation is a statistical value between zero and one that conveys the métier of the logistic association between a single predictor and response variables. A value approaching one indicates strong relationship and value approaching 0 denotes weak or no relationship.

We calculate the absolute value of the logistic correlation between all predictors and response variable as presented in Table 2. Since "diagnosis method" can directly predict the presence or absence of MM, we exclude it from all further analysis.

Table 3 shows the dependent variable coding. It tells us that "healthy" is coded as 0 and 1 represents "mesothelioma."

We then measure the performance of the baseline model. Baseline model is a model that does not include the explanatory variables.

Explanatory variables, also known as independent or predictor variables, are factors that are operationalized and used in a regression to predict a given outcome (Leech *et al*., 2015). The predictions of the baseline model are based purely on the frequency of occurrence of the category in the dataset.

Then we moved to the regression model that includes the explanatory variables and conducted The Omnibus tests. The Omnibus Tests of Model Coefficients measures the improvement of the new model over the baseline model (Leech *et al*., 2015). It uses chi-square tests to see if there is a significant difference between the Log-likelihoods of the baseline model and the new model (Leech *et al*., 2015). If the new model has a significantly reduced the Log-likelihood, then it suggests that the new model is elucidating more of the variance in the outcome and is an improvement. We calculated log likelihood and pseudo-R-square to determine the variation in the outcome. Consecutively, Hosmer and Lemeshow test (Leech *et al*., 2015) was conducted to determine the goodness on the model.

**Table 2:** Logistic correlation with the target variable

| Sl. No. | Input variable | Logistic correlation with "class of diagnosis." |
|---|---|---|
| 1 | Age | 0.066 |
| 2 | Gender | 0.155 |
| 3 | City | 0.029 |
| 4 | Asbestos exposure | 0.079 |
| 5 | Type of MM | 0.134 |
| 6 | Duration of asbestos exposure | 0.069 |
| 7 | Diagnosis method | 1.000 (excluded from analysis) |
| 8 | Keep side | 0.105 |
| 9 | Cytology | 0.029 |
| 10 | Duration of symptoms | 0.022 |
| 11 | Dyspnea | 0.026 |
| 12 | Ache on chest | 0.050 |
| 13 | Weakness | 0.060 |
| 14 | Habit of cigarette | 0.055 |
| 15 | Performance status | 0.039 |
| 16 | White blood | 0.050 |
| 17 | Cell count (WBC) | 0.052 |
| 18 | Hemoglobin (HGB) | 0.032 |
| 19 | Platelet count (PLT) | 0.065 |
| 20 | Sedimentation | 0.006 |
| 21 | Blood Lactic Dehydrogenize (LDH) | 0.014 |
| 22 | Alkaline Phosphate (ALP) | 0.041 |
| 23 | Total protein | 0.018 |
| 24 | Albumin | 0.041 |
| 25 | Glucose | 0.014 |
| 26 | Pleural lactic dehydrogenize | 0.036 |
| 27 | Pleural protein | 0.035 |
| 28 | Pleural albumin | 0.071 |
| 29 | Pleural glucose | 0.016 |
| 30 | Dead or not | 0.039 |
| 31 | Pleural effusion | 0.031 |
| 32 | Pleural thickness on tomography | 0.011 |
| 33 | Pleural level of acidity (pH) | 0.041 |
| 34 | C reactive protein (CRP) | 0.118 |

**Table 3:** Dependent variable encoding

| Original Value | Internal value |
|---|---|
| Healthy | 0 |
| Mesothelioma | 1 |

**Table 4:** Association rule model setting

| | |
|---|---|
| Maximum Number of Rules | 10 |
| Minimum Condition Support | 0.05 |
| Minimum Confidence | 0.10 |
| Minimum Rule Support | 0.05 |
| Minimum Lift | 2.00 |
| Maximum Number of Items in a Rule | 10 |
| Maximum Number of Items in a Condition | 6 |
| Maximum number of Items in a Prediction | 3 |
| Use only True Value for Flag Fields | True |
| Allow Rules without Conditions | False |
| Evaluation Measure Sorting the Rules | Confidence |

*Performance Measures*

When evaluating supervised training results, it is essential to check the performance of the training and testing through the accuracy and the AUC values. The performance measures come from equations based on the contingency matrix. This matrix is based on verifying combinations of true and false cases. The cases are true positive TP, true negative TN, false negative FN and false positive FP (Choudhury, 2018). Accuracy shows the percentage of correctly estimated true positive cases in the dataset. Overall prediction accuracy (Lotfi and Keshavarz, 2014) was used to identify the best fit model.

*Association Rule*

An association rule is an implication expression of the form $X \rightarrow Y$, where $X$ and $Y$ are disjoint sets (McCormick and Salcedo, 2017). The strength of an association rule is measured concerning its support and confidence. Support ($s$) determines how often a rule applies to a given dataset, while confidence (*c*) determines how frequently items in *Y* appear in transactions that contains *X* (Leech *et al*., 2015):

$$s(X \rightarrow Y) = (\sigma(X \cup Y))/N \quad (2)$$

$$c(X \rightarrow Y) = (\sigma(X \cup Y))/(\sigma(X)) \quad (3)$$

Previously unknown associations in the medical domain have been identified with the help of association rule mining in the medical literature (Leech *et al*., 2015). Association rule mining has been used to find disease-disease, disease-finding and disease-drug co-occurrences in electronic health record data (Hristovski *et al*., 2001; Swanson, 1990). We investigate the factors which contribute to MM disease. In our study, association rule mining, a computational intelligence approach, is used to identify these factors.

Table 4 shows the association model used in this study.

## Results

### Logistic Regression

Table 5 shows the processing summary of the training dataset. It tells us that the analysis includes 220 instances and no patients have missing data.

Table 6 describes the baseline model. In this study, the baseline model always guesses '0' because most participants were not affected by MM. The overall percentage row shows that this approach to prediction is correct 72.3% 0 of the time.

Table 7 shows us the coefficient for the constant ($B_0$). According to this table, the base model with just the constant is a statistically significant predictor of the outcome (p<0.001).

Table 8 shows the omnibus tests outcome. It has three rows: (a) step, (b) block and (c) model. The Model row compares the new model to the baseline. The Step and Block rows are only essential if explanatory variables were added to the model in a stepwise or hierarchical manner. If we were building the model up in stages, then these rows would compare the Log-likelihoods of the newest model with the previous version to ascertain whether or not each new set of explanatory variables were triggering improvements. In our study, we have added all explanatory variables in one block and therefore have only one step.

The Sig. values are p<0.001, which indicates the accuracy of the model improves after adding the explanatory variables. Table 9 provides the log likelihood and pseudo-R2 values for the full model. The R2 values show the approximate variation in the outcome. We prefer to use the Nagelkerke's R2 which suggests that the model explains 43.50% of the variation in the outcome.

**Table 5:** Case processing summary

| Unweighted cases | | N | Percent |
|---|---|---|---|
| Selected cases | Included in analysis | 220 | 100.00% |
| | Missing cases | 0 | 0 |
| | Total | 220 | 100.00% |
| Unselected cases | | 0 | 0.0 |
| Total | | 220 | 100.00% |

**Table 6:** Classification table of the baseline model

| | | | Predicted | | |
|---|---|---|---|---|---|
| | | | Class_of_diagnosis | | |
| | Observed | | 0.0 | 1.0 | Percentage correct |
| Step 0 | class_of_diagnosis | 0.0 | 159 | 0 | 100.00% |
| | | 1.0 | 61 | 0 | 0 |
| | Overall Percentage | | | | 72.30% |

**Table 7:** Variables in the equation

| | | B | S.E. | Wald | df | Sig. | Exp(B) |
|---|---|---|---|---|---|---|---|
| Step 0 | Constant | -0.958 | 0.151 | 40.463 | 1 | 0.000 | 0.384 |

**Table 8:** Omnibus test of model coefficients

| | | Chi-square | df | Sig. |
|---|---|---|---|---|
| Step 1 | Step | 78.919 | 43 | 0.001 |
| | Block | 78.919 | 43 | 0.001 |
| | Model | 78.919 | 43 | 0.001 |

**Table 9:** Model summary

| Step | -2 Log likelihood | Cox and Snell R Square | Nagelkerke R Square |
|---|---|---|---|
| 1 | 180.838[a] | 0.301 | 0.435 |

[a]Estimation terminated at iteration number 8 because log likelihood decreased by less than 0.001 percent

**Table 10:** Hosmer and Lemeshow test

| Step | Chi-square | df | Sig. |
|---|---|---|---|
| 1 | 6.312 | 8 | 0.612 |

**Table 11:** Classification table of the new model (training accuracy)

| | Observed | | Predicted class_of_diagnosis 0.0 | 1.0 | Percentage correct |
|---|---|---|---|---|---|
| Step 1 | class_of_diagnosis | 0.0 | 150 | 9 | 94.30% |
| | | 1.0 | 32 | 29 | 47.50% |
| | Overall Percentage training | | | | 81.40% |

**Table 12:** Comparing training and testing classification accuracy

| Partition | Training | | Testing | |
|---|---|---|---|---|
| Correct | 179 | **81.36%** | 66 | **63.46%** |
| Wrong | 41 | 18.64% | 38 | 36.54% |
| Total | 220 | | 104 | |

**Table 13:** Evaluation matrix

| Partition | Training | | Testing | |
|---|---|---|---|---|
| Model | AUC | GINI | AUC | GINI |
| Target | 0.844 | 0.688 | 0.613 | 0.277 |

**Table 14:** Rule statistics[a,b]

| Measurements | Minimum | Maximum | Mean | Standard deviation |
|---|---|---|---|---|
| Condition Support (%) | 8.33 | 10.19 | 9.20 | 0.75 |
| Confidence (%) | 62.50 | 70.37 | 66.67 | 3.23 |
| Rule Support (%) | 5.86 | 6.48 | 6.11 | 0.24 |
| Lift | 2.11 | 2.38 | 2.25 | 0.11 |
| Deployability (%) | 2.47 | 3.70 | 3.09 | 0.54 |

a. Number of Rules is 10
b. Number of Valid Events Data Source Records is 220

**Table 15:** Top five association rules sorted by confidence

| Condition | Pred. | Sorted by confidence (%) | Condition Support (%) | Rule Support (%) | Lift | Deployability (%) |
|---|---|---|---|---|---|---|
| dyspnoea weakness 7.600 ≤ cell count (WBC) < 11.200 glucose ≤ 132.200 1.760 ≤ pleural albumin < 2.640 | Mesothelioma | 70.37 | 8.33 | 5.86 | 2.38 | 2.47 |
| asbestos exposure dyspnoea weakness 7.600 ≤ cell count (WBC) < 11.200 1.760 ≤ pleural albumin < 2.640 | Mesothelioma | 69.23 | 8.02 | 5.56 | 2.34 | 2.47 |
| dyspnoea weakness 7.600 ≤ cell count (WBC) < 11.200 1.760 ≤ pleural albumin < 2.640 | Mesothelioma | 68.97 | 8.95 | 6.17 | 2.33 | 2.78 |
| dyspnoea weakness 7.600 ≤ cell count (WBC) < 11.200 1.760 ≤ pleural albumin < 2.640 dead or not | Mesothelioma | 68.97 | 8.95 | 6.17 | 2.33 | 2.78 |
| dyspnoea weakness 7.600 ≤ cell count (WBC) < 11.200 1.760 ≤ pleural albumin < 2.640 pleural effusion | Mesothelioma | 67.86 | 8.64 | 5.86 | 2.29 | 2.78 |

As shown in Table 10, the Hosmer and Lemeshow test of the goodness of fit suggests the model is an excellent fit to the data as p = 0.612 (>0.05). However, the chi-squared statistic on which it is based is dependent on sample size, so the value cannot be interpreted in isolation from the size of the sample.

The Table 11 shows the training accuracy of the model that includes all the explanatory variables. The new model is now correctly classifying the outcome for 81.40% of the cases compared to 72.30% in the baseline or null model. The classification accuracy for class 1 increased from zero to 47.50%.

Table 12 compares the testing and training accuracy of the logistic regression model. Logistic regression gives a classification training accuracy of 81.36% and testing accuracy of 63.46%. The significant difference between the training and testing accuracy is due the small sample size and data bias.

Table 13 compares the AUC and GINI of training and testing model.

*Association Rule*

Table 14 shows the association rule statistics.

The following Table 15 shows the top 5 association rules and their evaluation statistics.

The association rule shows that, "dyspnea", "weakness", "WBC count" and "pleural albumin" indicates the presence of Mesothelioma.

# Discussion

MPM is a belligerent cancer that is arduous to diagnose, specifically at early stages. In most of the developing nations, the reserves for histopathological diagnosis of suspected cases are limited. MPM is also a sporadic disease first diagnosed in 1962 (Musk *et al.*, 2011), with a very low likelihood of an individual suffering from this type of cancer.

A study exhibits the survival experience of Australian individuals with MM and found that survival has improved for each decade from the 1960s through 2000s (Musk *et al.*, 2011). The progressive advances in survival time was observed was plausibly due to enhanced prognosis with time which resulted in earlier presentation, diagnosis and improved treatment (Musk *et al.*, 2011). A review of a rift of cases from the first and last decades of 1960s and 2000s respectively, showed that, although the time between first presentation and diagnosis of MM did not alter (Musk *et al.*, 2011); rather, time between reported onset of symptoms and diagnosis did reduce significantly (63 days in the 1970s to 31 days in the 2000s). This insinuates that earlier diagnosis is merely due to patients' awareness of their symptoms, leading to earlier presentation in primary care or speedier referral by general practitioners to specialists. It is also possible that the improvement in diagnosing and treating MM in the early 90s burgeoned due to advancement in medical science or bias in reporting clinical trials and outcome.

In recent years, subjects suffering from the sarcomatoid subtype of MM were excluded from published clinical trials of active treatment because they did not respond as well as those with epithelioid or biphasic subtypes (Musk *et al.*, 2011). Barring of these and older patients from clinical trials tends to bias the overall survival of any study's participants giving a wrong impression that the prognosis of MM is improving.

Longer survival times for females have been reported previously (Kanazawa *et al.*, 2006; Marinaccio *et al.*, 2007; Mirabelli *et al.*, 2009; Neumann *et al.*, 2004); However, the biological cause responsible are unknown. Some studies have proposed that the difference may be due to misclassification as peritoneal MM of other abdominal neoplasms in females (Marinaccio *et al.*, 2007; Mirabelli *et al.*, 2009) Italian National Mesothelioma Register (Marinaccio *et al.*, 2007) reported augmented survival times in females with peritoneal, but not pleural.

Despite increasing resources and treatment expenses of MM there have been only modest improvements in survival over the past 40 yrs. Population-based study shows that median survival overall is still limited to less than a year from the time of diagnosis.

Therefore, primary diagnosis and prevention remains the most urgent priority for MM. Predictive analytics has the potential to advocate primary diagnosis of MM, increase the likelihood to patient survival.

# Conclusion

In this study we provide a prediction model using logistic regression to diagnose the presence of MM based on early stage symptoms.

We can infer from the results that logistic regression can improve primary diagnosis of mesothelioma disease and is a better approach than using no predictive model. The underfitting (high training accuracy and low testing accuracy) behavior of the logistic model was also observed during this study. This study identifies, "dyspnea," "weakness," "WBC count," and "pleural albumin" as the essential attributes that indicates the presence of mesothelioma disease.

However, our study is of limited statistical power to estimate sound results due to small sample size. Another, more general, limitation is that we cannot estimate the effect of age and gender as risk factors of developing an MPM.

# Availability of Data and Material

All data analyzed during this study are included in this published article and its supplementary information files.

# Funding Information

This study was not funded by any internal or external source.

## Ethics

No formal ethics approval was required in this particular case. The dataset used is publicly available for all researchers and no human participation was required for this study. The authors declare that they have no competing interests.

## References

Bueno, R., E. Stawiski, L.D. Goldstein, S. Durinck and A. De Rienzo *et al*., 2016. Comprehensive genomic analysis of malignant pleural mesothelioma identifies recurrent mutations, gene fusions and splicing alterations. Nat. Genet., 48: 407-416. DOI: 10.1038/ng.3520

Choudhury, A. and C.M. Greene, 2018. Prognosticating autism spectrum disorder using artificial neural network: Levenberg-marquardt algorithm. A. Clin. and Biomedical Research., 2: 184-193. DOI: 10.26502/acbr.50170058

Choudhury, A. and E. Khan, 2018. Decision support system for renal transplantation. Proceedings of the IISE Annual Conference, (AC' 18), IISE, Orlando.

Choudhury, A. and Wesabi, 2018. Classification of cervical cancer dataset. Proceedings of the IISE Annual Conference, (AC' 18), IISE, Orlando.

Choudhury, A., 2018. Evaluating patient readmission risk: A predictive analytics approach. Am. J. Eng. Applied Sci.

Frauenfelder, T., M. Tutic, W. Weder, R.P. Götti and R.A. Stahel *et al*., 2011. Volumetry: An alternative to assess therapy response for malignant pleural mesothelioma? Eur. Respir. J., 38: 162-168. DOI: 10.1183/09031936.00146110

Gill, R., D. Naidich, A. Mitchell, M. Ginsberg and J. Erasmus *et al*., 2016. North American multicenter volumetric ct study for clinical staging of malignant pleural mesothelioma: Feasibility and logistics of setting up a quantitative imaging study. J. Thorac. Oncol., 11: 1335-1344. DOI: 10.1016/j.jtho.2016.04.027

Hristovski, D., J. Stare, B. Peterlin and S. Dzeroski, 2001. Supporting discovery in medicine by association rule mining in Medline and UMLS. Stud Health Technol. Inform., 84: 1344-1348. PMID: 11604946

Kanazawa, N., A. Ioka, H. Tsukuma, W. Ajiki and A. Oshima, 2006. Incidence and survival of mesothelioma in Osaka, Japan. Japanese J. Clin. Oncol., 36: 254-257. DOI: 10.1093/jjco/hyl018

Leech, N.L., K.C. Barrett and G.A. Morgan, 2015. IBM SPSS for Intermediate Statistics. 5th Edn., Routledge, New York and London, ISBN-10: 1136334947, pp: 368.

Leigh, J., C. Corvalan, A. Grimwood, G. Berry and D. Ferguson *et al*., 1991. The incidence of malignant mesothelioma in Australia 1982-1988. Am. J. Ind. Med., 20: 643-655. DOI: 10.1002/ajim.4700200507

Lotfi, E. and A. Keshavarz, 2014. Gene expression microarray classification using PCA-BEL. Comput. Biol. Med., 180-187: 180-187. DOI: 10.1016/j.compbiomed.2014.09.008

Marinaccio, A., A. Binazzi, G. Cauzillo, D. Cavone and R. De Zotti *et al*., 2007. Analysis of latency time and its determinants in asbestos related malignant mesothelioma cases of the Italian register. Eur. J. Cancer, 43: 2722-2728. DOI: 10.1016/j.ejca.2007.09.018

McCormick, K. and J. Salcedo, 2017. Regression with Categorical Outcome Variables. In: SPSS Statistics for Data Analysis and Visualization, McCormick, K. and J. Salcedo (Eds.), Wiley, Indianapolis, ISBN-10: 1119003555, pp: 71-93.

McDonald, J.C. and A.D. McDonald, 1996. The epidemiology of mesothelioma in historical context. Eur. Respiratory J., 9: 1932-1942. DOI: 10.1183/09031936.96.09091932

Metintas, M., S. Metintas, G. Ak, S. Erginel and F. Alatas *et al*., 2008. Epidemiology of pleural mesothelioma in a population with non-occupational asbestos exposure. Respirology, 13: 117-121. 10.1111/j.1440-1843.2007.01187.x

Mirabelli, D., S. Roberti, M. Gangemi, R. Rosato and F. Ricceri *et al*., 2009. Survival of peritoneal malignant mesothelioma in Italy: A population-based study. Int. J. Cancer, 124: 194-200 DOI: 10.1002/ijc.23866

MN, 2018. Mesothelioma News. http://www.mesotheliomanews.com/medical/mesothelioma-diagnosis/pleural-mesothelioma/

Musk, A.W., N. Olsen, H. Alfonso, A. Reid and R. Mina *et al*., 2011. Predicting survival in malignant mesothelioma. Eur. Respiratory J., 38: 1420-1424. DOI: 10.1183/09031936.00000811

Neumann, V., A. Rütten, M. Scharmach, K.M. Müller and M. Fischer, 2004. Factors influencing long-term survival in mesothelioma patients - Results of the German mesothelioma register. Int. Arch. Occupat. Environ. Health, 77: 191-199. DOI: 10.1007/s00420-003-0498-6

Parodi, S., R. Filiberti, P. Marroni, R. Libener and G.P. Ivaldi, 2015. Differential diagnosis of pleural mesothelioma using Logic Learning Machine. BMC Bioinform., 16: 1471-2105. DOI: 10.1186/1471-2105-16-S9-S3

Pass, H., D. Giroux, C. Kennedy, E. Ruffini and A.K. Cangir *et al*., 2016. The IASLC mesothelioma staging project: Improving staging of a rare disease through international participation. J. Thorac. Oncol., 11: 2082-2088.
DOI: 10.1016/j.jtho.2016.09.123

Peto, J., J. Hodgson, K. Matthews and J. Jones, 1995. Continuing increase in mesothelioma mortality in Britain. Lancet, 345: 535-539.
DOI: 10.1016/S0140-6736(95)90462-X

Spirtas, R., G. Beebe, R. Connelly, W. Wright and J. Peters *et al*., 1986. Recent trends in mesothelioma incidence in the United States. Am. J. Ind. Med, 9: 397-407. DOI: 10.1002/ajim.4700090502

Swanson, D.R., 1990. Medical literature as a potential source of new knowledge. Bull. Med. Libr. Assoc., 78: 29-37. PMID: 2403828

Tin, K.O., 1995. Random decision forests. Proceedings of the 3rd International Conference on Document Analysis and Recognition, Aug. 14-16, IEEE Xplore Press, Montreal, Quebec, Canada, pp: 278-282.
DOI: 10.1109/ICDAR.1995.598994

Vogelzang, N., J. Rusthoven, J. Symanowski, C. Denham and E. Kaukel *et al*., 2003. Phase III study of pemetrexed in combination with cisplatin versus cisplatin alone in patients with malignant pleural mesothelioma. J. Clin. Oncol., 21: 2636-2644.
DOI: 10.1200/JCO.2003.11.136

Zalcman, G., J. Mazieres, J. Margery, L. Greillier and C. Audigier-Valette *et al*., 2016. Bevacizumab for newly diagnosed pleural mesothelioma in the Mesothelioma Avastin Cisplatin Pemetrexed Study (MAPS): A randomized, controlled, open-label, phase 3 trial. Lancet, 387: 1405-1414.
DOI: 10.1016/S0140-6736(15)01238-6